\documentclass[serif,twocolumn]{wiley-article}

\usepackage{siunitx}
\usepackage{hyperref}
\hypersetup{
	colorlinks=true,
	linkcolor=black,
	filecolor=black,      
	urlcolor=blue,
	citecolor=black,
	}
\usepackage[sorting=none,maxnames=2,minnames=1,maxbibnames=99,style=numeric-comp,giveninits=true]{biblatex} 
\bibliography{sample.bib}
\usepackage{cleveref}
\usepackage[version=4]{mhchem} 
\usepackage{comment}

\papertype{Original Article}
\title{Ultralow-cost magnetocaloric compound for cryogenic cooling}

\author[1\authfn{1}]{Wei Liu}
\author[1]{Benjamin Theisel}
\author[1]{Yulia Klunnikova}
\author[1]{Konstantin Skokov}
\author[1]{Oliver Gutfleisch}

\affil[1]{TU Darmstadt, Institute of Materials Science, 64287 Darmstadt, Germany}

\corraddress{Wei Liu}
\corremail{\authfn{1} wei.liu@tu-darmstadt.de}

\fundinginfo{The Clean Hydrogen Partnership within the framework of the HyLICAL project (Grant No. 101101461) \\
The Deutsche Forschungsgemeinschaft (DFG) within the CRC/TRR 270 (Project-ID 405553726)}

\runningauthor{Wei LIU et al.}

\begin{document}

\begin{frontmatter}
\maketitle

\begin{abstract}
Cost-effective materials are essential for large-scale deployment. The emerging magnetocaloric hydrogen liquefaction technology could transform the liquid hydrogen industry due to its potential in achieving higher efficiency. Most studies of the cryogenic magnetocaloric effect (MCE) have focused on resource-critical rare-earth-based compounds. Here we report on an ionic magnetocaloric compound \ce{FeCl2} which is based on ultralow-cost elements, as a candidate working material for hydrogen liquefaction. \ce{FeCl2} shows both inverse and conventional MCE. From 0 to \qty{1.5}{\tesla}, the inverse effect yields a positive magnetic entropy change ($\Delta S_T$) of about \qty{5}{\joule\per\kelvin\per\kilogram} near \qty{20}{\kelvin}, then declines toward zero at higher fields. In contrast, the conventional (negative) response strengthens with field. The $\Delta S_T$ reaches \qty{18.6}{\joule\per\kelvin\per\kilogram} near \qty{20}{\kelvin} in magnetic fields of \qty{5}{\tesla}. This value exceeds most light rare-earth-based compounds and approaches that of heavy rare-earth-based compounds. In magnetic fields of \qty{5}{\tesla}, the adiabatic temperature change reaches about \qty{3.6}{\kelvin}. The large $\Delta S_T$, along with the low cost of the elements in \ce{FeCl2}, are prerequisites for inexpensive industrial-scale production, giving the prospect of a practical magnetocaloric candidate for hydrogen liquefaction in the 20 $\sim$ \qty{77}{\kelvin} temperature window.

% Please include a maximum of seven keywords
\keywords{Magnetism, magnetocaloric effect, hydrogen liquefaction, antiferromagnetism, ionic compound}
\end{abstract}

\end{frontmatter}

\begin{figure*}[hb!]
    \centering
    \includegraphics[width=\linewidth]{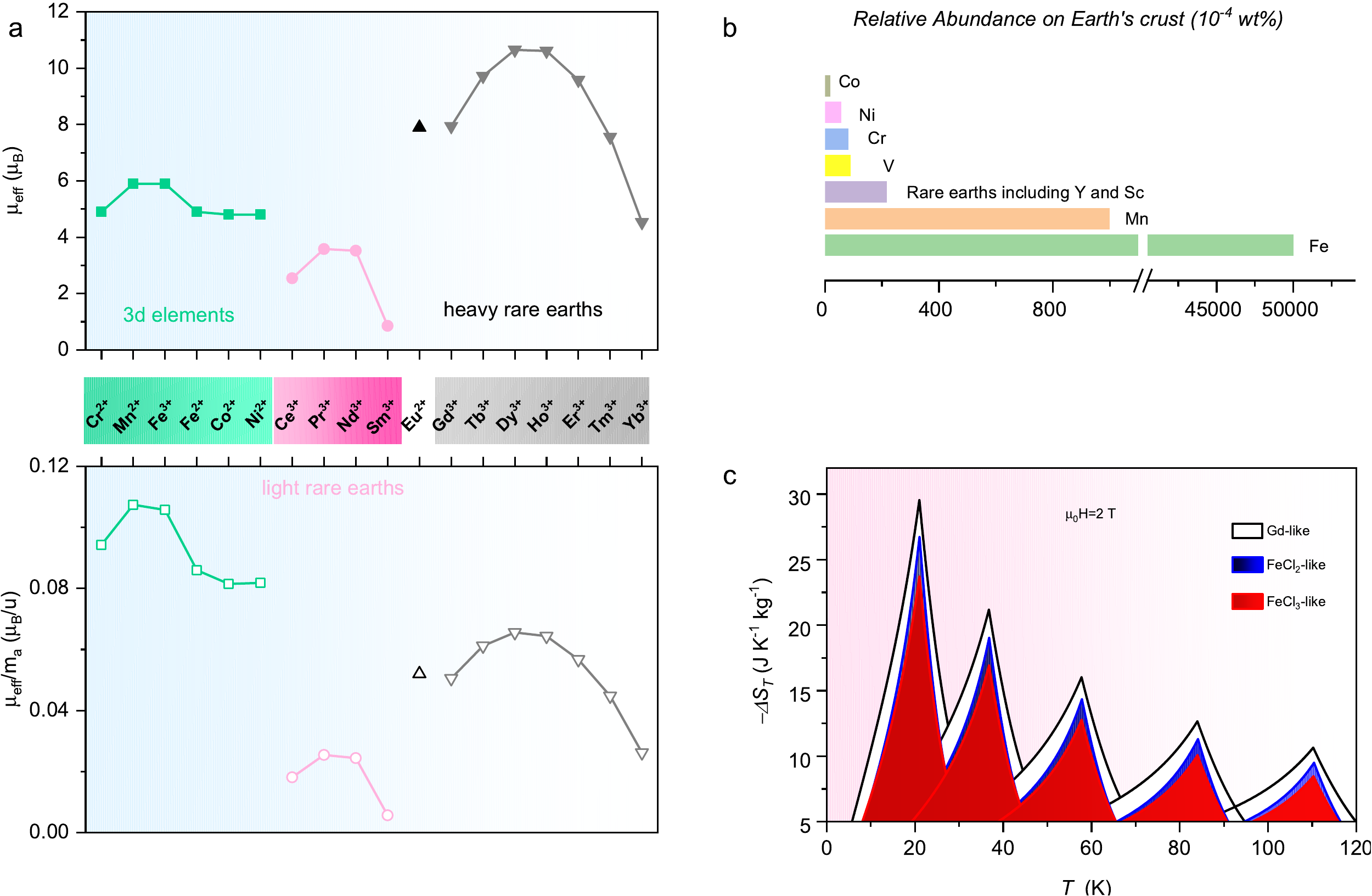}
    \caption{(a) Effective magnetic moments of the ions of transition metals, light rare earths, Eu, and heavy rare earths and their Effective magnetic moment ($\mu_{eff}$) per atomic mass ($m_a$). (b) Relative abundance of Co, Ni, Cr, V, rare earths including Y and Sc, Mn, and Fe in the earth's crust. (c) $\Delta S_T$ of Gd-like, \ce{FeX2}, and \ce{FeX3} in magnetic fields of 2 T obtained from a mean-field approach.}
    \label{Figure1}
\end{figure*}

Magnetocaloric hydrogen liquefaction has the potential to achieve higher efficiency than conventional technologies based on Joule--Thomson expansion, opening a path to affordable liquid hydrogen by reducing energy consumption during the liquefaction process \cite{liu_matter_2024,liu_study_2022,Franco.2018,zhang_advanced_2024,romero-muniz_magnetocaloric_2023,Cirillo2025,Ma2025,Zheng2017,Czernuszewicz2025}. If hydrogen gas is pre-cooled by liquid nitrogen to approximately \qty{77}{\kelvin} (the condensation point of nitrogen), magnetocaloric hydrogen liquefaction requires magnetocaloric materials that operate efficiently in the temperature range from \qty{77}{\kelvin} down to about \qty{20}{\kelvin} (the condensation point of hydrogen) \cite{Matsumoto2011,Park2017,Tang2022}. Developing magnetocaloric materials that exhibit large magnetic entropy and adiabatic temperature changes within this temperature window under affordable magnetic fields (e.g., 5 T generated by a superconducting magnet \cite{Barclay2019,kamiya_active_2022,Kamiya2025}) is therefore of fundamental importance for realizing practical magnetocaloric hydrogen liquefaction \cite{liu_matter_2024}. Rare-earth-based compounds have been studied extensively because their ions carry large magnetic moments and yield strong cryogenic magnetocaloric responses \cite{liu_matter_2024,liu_study_2022,Franco.2018}. However, rare-earth elements are among the most critical raw materials \cite{liu2023designing,romero-muniz_magnetocaloric_2023,Lai2022,Levinsky2024Co4OH6SO4enH2NatComm}, raising concerns about the scalability of rare-earth-based solutions for large-scale hydrogen liquefaction.

Compared with the more extensively studied rare-earth systems, 3d–metal–based materials remain relatively underexplored for magnetocaloric hydrogen liquefaction \cite{TishinSpichkin2016,PecharskyGschneidner1999}. However, considering their significant magnetic moments, 3d–metal–based compounds should also have strong potential for large magnetocaloric effects. Near room temperature, many 3d-metal–based materials—such as the \ce{Fe2P}-type compounds \cite{Tegus2002}, the NiMn-based heuslers \cite{Liu2012,krenke_inverse_2005,wei_realization_2015,Wei2016}, and \ce{MnAs} \cite{Wada2001,Gutfleisch.2011,Gutfleisch.2016}-exhibit a pronounced magnetocaloric effect. As shown in \Cref{Figure1} (a), 3d–metal ions carry substantial magnetic moments; although these are generally smaller than those of the heavy rare-earth ions \cite{Coey2010}, several high-spin trivalent 3d ions (e.g., \ce{Cr^{3+}}, \ce{Mn^{3+}}, \ce{Fe^{3+}}) match or exceed those of most light rare-earth ions. Moreover, on a mass-normalized basis shown in \Cref{Figure1} (a), \ce{Mn^{3+}} and \ce{Fe^{3+}} exhibit magnetic moments that surpass even \ce{Ho^{3+}}, the rare-earth ion with one of the largest per-atom moments \cite{CRC100}.

In many intermetallics, $3d$ electrons are largely itinerant, and the magnetic moments of $3d$ ions are consequently reduced from their free-ion values by band formation and hybridization \cite{RhodesWohlfarth1963,Moriya1985}. By contrast, ionic compounds typically lack itinerant carriers, leading to localized $3d$ moments \cite{Zaanen1985,Imada1998}. From this perspective, the magnetic moments in ionic compounds can approach those of the corresponding $3d$ ions, although crystal-field splitting, covalency, and low-spin configurations can reduce the observed moments relative to the high-spin limit \cite{AbragamBleaney1970}.

Among the 3d transition metals with large magnetic moments, iron (\ce{Fe}) stands out due to its abundance and low cost~\cite{EU_CRM_2023,DOE_Critical_2023}. \Cref{Figure1} (b) shows the relative abundances of \ce{Fe} and the rest of the elements listed in \Cref{Figure1} (a) (data is taken from Ref.\cite{Yaroshevsky2006}). Notably, \ce{Fe} is the most abundant, surpassing the combined abundance of all others. This makes Fe-based magnetocaloric materials particularly attractive for hydrogen liquefaction from a materials-criticality perspective, with \ce{La(Fe,Si)13} as a prominent example for a broad range of temperatures from room temperature \cite{Fujita2003,Hu2002,Krautz2014} down to cryogenic temperatures \cite{Lai2021,Beckmann2024,Strassheim2025}.

To gain an intuitive understanding of the potential magnetic entropy change ($\Delta S_T$) in Fe-based materials, we consider three simplified ferromagnetic model systems: a Gd-like system, an \ce{FeCl2}-like system, and an \ce{FeCl3}-like system.

The magnetic entropy $S_m$ of these toy models is given by~\cite{Tishin.2003}:
\begin{equation} 
    S_m = N_M k_B \left [ \ln \left ( \frac{\sinh \left ( \tfrac{2J+1}{2J}y \right ) }{\sinh \left ( \tfrac{1}{2J}y \right )} \right ) - y B_J(y) \right ] ,
\end{equation}
where $N_M$ is the number of magnetic atoms, $k_B$ is the Boltzmann constant, $J$ is the total angular momentum quantum number, $B_J(y)$ is the Brillouin function, and 
\begin{equation}
       y = \frac{g_J J \mu_B \mu_0 H + \tfrac{3J}{J+1} k_B T_C B_J(y)}{k_B T} \, .
\end{equation}  

The $\Delta S_T$ values of the three model systems are obtained by taking $J$ from Gd for the Gd-like system, from \ce{Fe^{2+}} for the \ce{FeCl2}-like system, and from \ce{Fe^{3+}} for the \ce{FeCl3}-like system. As shown in \Cref{Figure1}(b), both the \ce{FeCl2}-like and \ce{FeCl3}-like systems can exhibit $\Delta S_T$ values that are large and comparable to those of the Gd-like system, particularly near \qty{20}{\kelvin}. This mean-field analysis highlights the potential of Fe-based ionic compounds to achieve significant magnetocaloric effects.

Here, we investigate the magnetocaloric effect (MCE) of \ce{FeCl2}. This material undergoes an antiferromagnetic transition at approximately \qty{24}{\kelvin} and exhibits a field-induced spin-flip (metamagnetic) transition with a critical field around \qty{1.5}{\tesla}, implying a substantial magnetic entropy change ($\Delta S_T$) for fields above \qty{2}{\tesla} \cite{Xu_thermalHall_2023,Jacobs_PR_1967}. \ce{FeCl2} powders were obtained from Merck/MilliporeSigma with a purity of 98\%, and magnetization measurements were performed using a Quantum Design MPMS3 SQUID magnetometer. To measure the heat capacity, the \ce{FeCl2} powder was first uniaxially pressed into a $\phi$ 15 mm pellet inside an Ar-filled glovebox. The pellet was then sealed in a rubber bag and subjected to cold isostatic pressing at 0.56 GPa to obtain a highly densified block. A small piece was taken from the block for the heat capacity measurement in a PPMS (physical property measurement system) from Quantum Design using the 2$\tau$ method.

\begin{figure*}[ht!]
    \centering
    \includegraphics[width=0.8\linewidth]{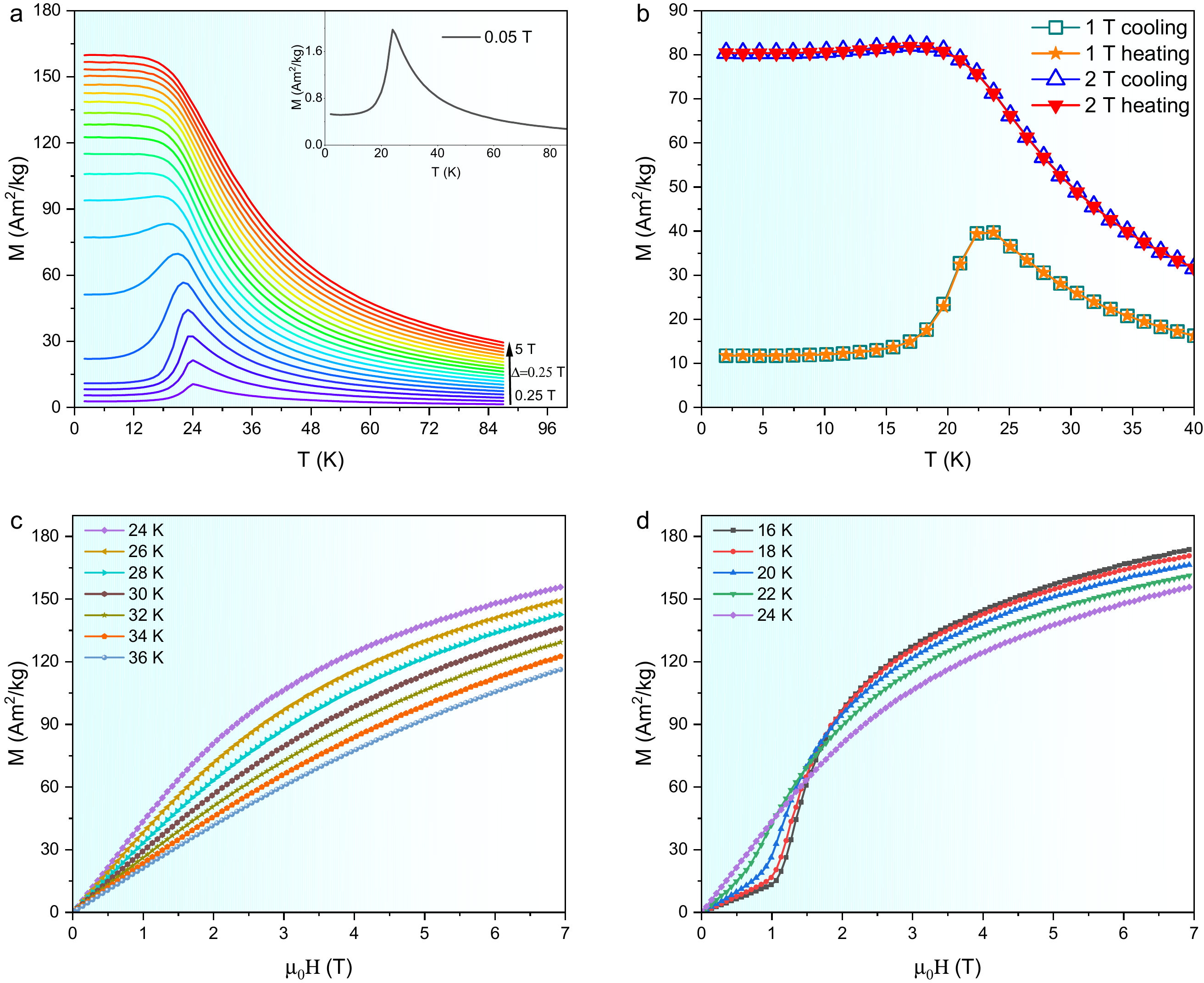}
    \caption{(a) Magnetization as a function of temperature up to \qty{5}{\tesla} with a step of \qty{0.25}{\tesla}. The inset shows the case in \qty{0.05}{\tesla}. (b) Magnetization as a function of temperature in magnetic fields of 1 and \qty{2}{\tesla} with a ramping rate of \qty{0.5}{\kelvin\per\min} and stabilized temperatures before each measurement. (c) Magnetization as a function of temperature before the antiferromagnetic transition. (d) Magnetization as a function of temperature across the antiferromagnetic transition. }
    \label{fig2}
\end{figure*}

\begin{figure*}[ht!]
    \centering
    \includegraphics[width=0.9\linewidth]{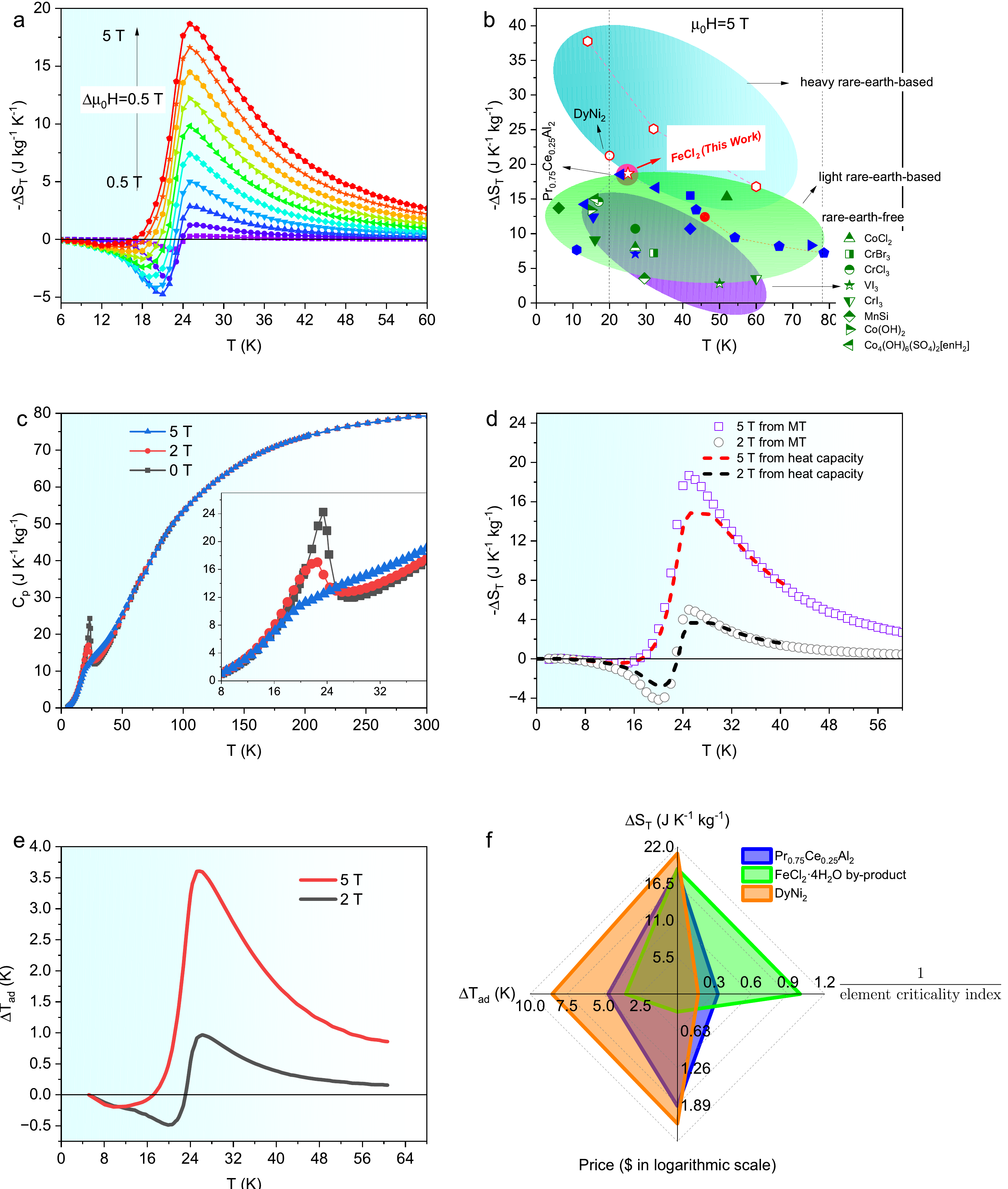}
    \caption{(a) $\Delta S_T$ of \ce{FeCl2} in magnetic fields up to \qty{5}{\tesla}. (b) Comparing $\Delta S_T$ of \ce{FeCl2} with heavy rare-earth-based \cite{liu_study_2022}, light rare-earth-based \cite{liu2023designing,zhang_magnetic_2015,dong_effect_2019,wang_magnetic_2014,zheng_nearly_2014,zheng_large_2018,Wang.2014,Paixao.2020,Plaza.2009,Li.2011c,Li.2009,Ma.2019b,Lyu.2020, liu_study_2022,kumar_magnetism_2008}, and non-rare-earth-based magnetocaloric materials for cryogenic cooling in $\mu_0\Delta H=\qty{5}{\tesla}$, including \ce{CrCl3}~\cite{Liu2020CrCl3PRB}, \ce{CrBr3}~\cite{Yu2019CrBr3FrontiersPhys}, \ce{CrI3}~\cite{Liu2018CrI3PRB}, \ce{CoCl2}~\cite{Liu2010CoCl2JAC}, and $\beta$-\ce{Co(OH)2}~\cite{Liu2014BetaCoOH2MR}. (c) Heat capacity in magnetic fields of 0, 2, 5 T. The inset shows the area around the phase transition. (d) $\Delta S_T$ from heat capacity measurement in magnetic fields of 2 and 5 T in comparison with that obtained from magnetization measurements. (e) $\Delta T_{ad}$ from heat capacity measurement in magnetic fields of 2 and 5 T. (f) Comparison of $\Delta S_T$ at 5 T, elemental criticality, and material price for \ce{FeCl2} (non–rare-earth based), \ce{DyNi2} (heavy rare-earth based), and \ce{Pr_{0.75}Ce_{0.25}Al2} (light rare-earth based) near \qty{20}{\kelvin}. \ce{FeCl2} is assumed to be produced by dehydration of \ce{FeCl2*4H2O} above \qty{373}{\kelvin}. Prices of \ce{DyNi2} and \ce{Pr_{0.75}Ce_{0.25}Al2} were estimated by summing the elemental prices from Websites \cite{dy_ni_prices_2025,pr_ce_al_prices_2025}, while the price of \ce{FeCl2*4H2O} was obtained from Alibaba \cite{alibaba_fecl2_4h2o_2025}. Elemental criticality indices were estimated from Ref.~\cite{Gottschall.2019}.}
    \label{fig3}
\end{figure*}

\Cref{fig2} (a) shows the magnetization as a function of temperature of \ce{FeCl2} under a magnetic field up to \qty{5}{\tesla}. In magnetic fields below \qty{2}{\tesla}, the magnetization firstly increases with decreasing temperature but near \qty{24}{\kelvin} the magnetization decreases, indicating a transition to an antiferromagnetic state. With the magnetic field increasing, the magnetization drop becomes smaller and then vanishes above \qty{2}{\tesla}. This transition suggest a coexistence of inverse and conventional magnetocaloric effect. 

This phenomenon can be also seen from \Cref{fig2} (b) showing the magnetization as a function of magnetic fields up to \qty{7}{\tesla}. Above \qty{24}{\kelvin}, the MH curves below \qty{2}{\tesla} are almost linear, suggesting a paramagnetic state. In addition, the MH curves shift upward with decreasing temperature. This is a typical paramagnetic behavior where the magnetization increases towards lower temperature.  However, as shown in \Cref{fig2} (d), below \qty{24}{\kelvin} and \qty{2}{\tesla}, the MH curves shifts downwards with decreasing temperature. This is a typical behavior of paramagnetic-to-antiferromagnetic transition. It is worth mentioning that this compound exhibits a large magnetization above \qty{2}{\tesla} after the magnetic phase transition. In \qty{5}{\tesla} at \qty{2}{\kelvin} the magnetization is approaching \qty{160}{\ampere\square\meter\per\kilogram}. In addition, the magnetic transition of \ce{FeCl2} is hysteresis-free as the cooling and heating MT curves in 1 and \qty{2}{\tesla} perfectly overlap at a small temperature ramping rate of  \qty{0.5}{\kelvin\per\min}.

The magnetic entropy change, $\Delta S_T$, was obtained from the $M(T)$ curves measured with a field step of \qty{0.25}{\tesla} using the Maxwell relation \cite{moya_caloric_2014}
\begin{equation}
\Delta S_T = \mu_0 \int_0^H \left( \frac{\partial M}{\partial T} \right)_H dH \, ,
\end{equation}
where $\mu_0$ is the vacuum permeability. As shown in \Cref{fig3} (a), both the conventional and inverse magnetocaloric effects (MCE) coexist in \ce{FeCl2}. This is consistent with the magnetization data, which reveal a transition to an antiferromagnetic phase below about \qty{2}{\tesla}. The maximum positive $\Delta S_T$ associated with the inverse MCE reaches \qty{4.7}{\joule\per\kelvin\per\kilogram} at \qty{21}{\kelvin} for $\mu_0 H=\qty{1.5}{\tesla}$. For fields above \qty{1.5}{\tesla}, this positive $\Delta S_T$ decreases with increasing field and nearly vanishes by \qty{5}{\tesla}.

In contrast, below \qty{1}{\tesla} the conventional (negative) $\Delta S_T$ is modest, with a value of only \qty{1.2}{\joule\per\kelvin\per\kilogram} at \qty{1}{\tesla}. However, for $\mu_0 H>\qty{1}{\tesla}$, $|\Delta S_T|$ grows rapidly with field: at \qty{5}{\tesla} it reaches about \qty{5}{\joule\per\kelvin\per\kilogram} near \qty{5}{\kelvin}, and attains a large magnitude of approximately \qty{18.6}{\joule\per\kelvin\per\kilogram} at around \qty{24}{\kelvin}.

\Cref{fig3}(b) benchmarks the $\Delta S_T$ of \ce{FeCl2} against selected heavy–rare-earth, light–rare-earth, and non–rare-earth magnetocaloric materials at \qty{5}{\tesla}. Near \qty{20}{\kelvin} (close to the condensation point of hydrogen), \ce{FeCl2} surpasses all surveyed non–rare-earth compounds, including \ce{Co4(OH)6(SO4)2[enH2]}, which has been reported to exhibit a giant MCE. Polycrystalline \ce{FeCl2} delivers a $\Delta S_T$ comparable to that of single-crystal \ce{PrAlSi} and polycrystalline \ce{Pr_{0.75}Ce_{0.25}Al2}, which show the largest $\Delta S_T$ among light–rare-earth materials near \qty{20}{\kelvin}. Moreover, the $\Delta S_T$ of \ce{FeCl2} is even comparable to that of \ce{DyNi2}, a heavy–rare-earth magnetocaloric material well-known for its strong magnetocaloric performance at cryogenic temperature.

From the total entropy curves, $S(T, H)$, the magnetic entropy change ($\Delta S_T$) and adiabatic temperature change ($\Delta T_{ad}$) were evaluated according to
\begin{align}
\Delta S_T &= S(T, H) - S(T, 0),\, \\
\Delta T_{ad} &= T(S, H) - T(S, 0),
\end{align}
where $T(S, H)$ denotes the inverse function of $S(T, H)$ \cite{liu_study_2022}. The total entropy curve can be constructed via
\begin{equation}
    S(T, H) = \int_0^{H} \frac{C_p}{T}dT \, .
\end{equation}
The heat capacities measured under magnetic fields of 0, 2, and 5 T are presented in \Cref{fig3}(c). In magnetic fields of 0 and 2 T, the heat capacity anomalies exhibit a characteristic $\lambda$-type profile, indicative of a second-order phase transition. With increasing magnetic field, the peak height of the anomaly gradually decreases, and at 5 T the feature becomes indistinct.

The resulting $\Delta S_T$ and $\Delta T_{ad}$ are shown in \Cref{fig3}(d) and (e), respectively. $\Delta S_T$ derived from heat capacity shows a small deviation compared with that obtained from the loose powders. Such behavior likely originates from partial alignment of crystallites and the associated modification of magnetic anisotropy under compression. This interpretation is consistent with a suggestion from Dr. Hiroaki Mamiya (National Institute for Materials Science, Japan), who noted that simple orientation of \ce{FeCl2} powders can enhance the apparent magnetocaloric response \cite{Mamiya2023}. We gratefully acknowledge his kind communication and insightful comment regarding this effect. The maximum $\Delta T_{ad}$ of \ce{FeCl2} reaches approximately 1 K and 3.6 K under magnetic fields of 2 T and 5 T, respectively. 

In \Cref{fig3} (f) we compare $\Delta S_T$ and $\Delta T_{ad}$ in magnetic fields of \qty{5}{\tesla}, inverse values of element criticality index of Fe, Dy, and Pr, and prices of \ce{FeCl2} (non-rare-earth based), \ce{DyNi2} (heavy rare-earth based), and \ce{Pr_{0.75}Ce_{0.25}Al2} (light rare-earth based) with close to \qty{20}{\kelvin} assuming that \ce{FeCl2} is produced by heating \ce{FeCl2*4H2O} over \qty{373}{\kelvin}. These three have a similar $\Delta S_T$ range from about 18 to \qty{21}{\joule\per\kelvin\per\kilogram}. The \ce{FeCl2} shows dominated advantages over the rest two in terms of price and material criticality.

In conclusion, the ionic compound \ce{FeCl2} exhibits an excellent $\Delta S_T$ of about \qty{18.6}{\joule\per\kelvin\per\kilogram} at approximately \qty{24}{\kelvin} under \qty{5}{\tesla}, an affordable magnetic fields generated by superconducting magnets. This value exceeds that of most light–rare-earth magnetocaloric materials relevant to hydrogen liquefaction and rivals some heavy–rare-earth counterparts known for large cryogenic MCE. The $\Delta T_{ad}$ reaches about \qty{3.6}{\kelvin} in this field. Considering that \ce{FeCl2} contains only abundant, non-critical elements, and that \ce{FeCl2} can be produced by heating \ce{FeCl2*4H2O} which is a low-value by-product in many chemical industries, our work points toward a cost-effective pathway for magnetocaloric hydrogen liquefaction at scale.

\section*{Acknowledgment}
We gratefully acknowledge the related patent JP2023085746B2 and thank Dr. Hiroaki Mamiya (National Institute for Materials Science, Japan) for his kind communication regarding the patent and for his valuable suggestions on the improvement of the compressed \ce{FeCl2} pellets. This work was conceived and carried out independently.

We would also like to thank Dr. Wenjie Xie and Dr. Xingxing Xiao from the Materials and Resources Group, Department of Materials Science, TU Darmstadt, for their valuable support with the powder pressing procedures.

We gratefully acknowledge financial support from the Clean Hydrogen Partnership under the framework of the HyLICAL project (Grant No. 101101461) and from the Deutsche Forschungsgemeinschaft (DFG) through CRC/TRR 270 (Project-ID 405553726). 

\section*{Conflict of Interest}
The authors declare no competing financial interests. The authors note that they have communicated with Dr. Hiroaki Mamiya (NIMS, Japan) regarding the related patent JP2023085746B2; however, this study was conceived and carried out independently.

\printbibliography

\end{document}